# ECG-SMART-NET: A Deep Learning Architecture for Precise ECG Diagnosis of Occlusion Myocardial Infarction

Nathan T. Riek, Murat Akcakaya, Zeineb Bouzid, Tanmay Gokhale, Stephanie Helman, Karina Kraevsky-Philips, Rui Qi Ji, Ervin Sejdic, Jessica K. Zègre-Hemsey, Christian Martin-Gill, Clifton W. Callaway, Samir Saba, Salah Al-Zaiti

***Abstract*— Objective: In this paper we develop and evaluate ECG-SMART-NET for occlusion myocardial infarction (OMI) identification. OMI is a severe form of heart attack characterized by complete blockage of one or more coronary arteries requiring immediate referral for cardiac catheterization to restore blood flow to the heart. Two thirds of OMI cases are difficult to visually identify from a 12-lead electrocardiogram (ECG) and can be potentially fatal if not identified quickly. Previous works on this topic are scarce, and current state-of-the-art evidence suggests both feature-based random forests and convolutional neural networks (CNNs) are promising approaches to improve ECG detection of OMI. Methods: While the ResNet architecture has been adapted for use with ECG recordings, it is not ideally suited to capture informative temporal features within each lead and the spatial concordance or discordance across leads. We propose a clinically informed modification of the ResNet-18 architecture. The model first learns temporal features through temporal convolutional layers with 1xk kernels followed by a spatial convolutional layer, after the residual blocks, with 12x1 kernels to learn spatial features. Results: ECG-SMART-NET was benchmarked against the original ResNet-18 and other state-of-the-art models on a multisite real-word clinical dataset that consists of 10,393 ECGs from 7,397 unique patients (rate of OMI =7.2%). ECG-SMART-NET outperformed other models in the classification of OMI with a test AUC of 0.953 [0.921, 0.978]. Conclusion and Significance: ECG-SMART-NET can outperform the state-of-the-art random forest for OMI prediction and is better suited for this task than the original ResNet-18 architecture.**

## I. INTRODUCTION

Deep learning has been widely adopted for applications such as image classification, natural language processing, supply chain, and healthcare diagnostics [1][2][3][4]. Deeper models can learn more complex features in the data, however, when models get too deep, they start to decline in performance due to the vanishing gradient problem [5][6]. To overcome this problem, the ResNet architecture presents the use of skip connections [1]. ResNet was designed primarily for image classification tasks and the original architecture was trained on images with dimensions of 224 x 224 pixels. Since its conception, the ResNet architecture has been modified and used by many researchers for electrocardiogram (ECG) classification tasks [7][8][9]. Since ECGs are multi-channel time series data and not images, the original ResNet-18 architecture requires some architectural adjustments. In this work, we propose a modified ResNet-18 architecture, ECG-SMART-NET, to be used with 12-lead median beat ECGs for binary Occlusion Myocardial Infarction (OMI) classification.

OMI is a potentially fatal condition where blood flow is completely blocked in one or more coronary arteries resulting in acute myocardial infarction (i.e., heart attack). Thus, it is important that OMIs are quickly and correctly identified so that patients can receive life-saving reperfusion therapy to restore blood flow through blocked arteries and salvage myocardium. Unlike other heart conditions that can be more readily identified by ECG changes (i.e. ST-elevation myocardial infarction [STEMI]), OMI does not always present obvious patterns on an ECG recording, increasing risk of poor clinical outcomes for these patients from delays in reperfusion therapy [10]. This provides an opportunity for machine learning (ML) models to

This work was supported in part by the National Institutes of Health under Grant RO1HL137761, the National Center for Advancing Translational Sciences under grant KL2TR0011009, and the National Heart, Lung and Blood Institute under grant T32HL129964 *(Corresponding author: Nathan T. Riek).*

Nathan T. Riek (email: ntr14@pitt.edu) and Murat Akcakaya are with the Department of Electrical and Computer Engineering, University of Pittsburgh, Pittsburgh, PA, USA.

Zeineb Bouzid is with the School of Electrical and Computer Engineering, Georgia Institute of Technology, Atlanta, GA, USA.

Karina Kraevsky-Phillips, Tanmay Gokhale, Christian Martin-Gill, Clifton W. Callaway, and Samir Saba are with the University of Pittsburgh Medical Center (UPMC), Pittsburgh, PA, USA.

Tanmay Gokhale and Samir Saba are with the Division of Cardiology, University of Pittsburgh, Pittsburgh, PA, USA.

Karina Kraevsky-Phillips is with the Department of Acute and Tertiary Care Nursing, University of Pittsburgh, Pittsburgh, PA, USA.

Stephanie Helman is with the School of Medicine, Department of Medicine, Division of General Internal Medicine, University of Pittsburgh, Pittsburgh, PA, USA.

Christian Martin-Gill and Clifton W. Callaway are with the Department of Emergency Medicine, University of Pittsburgh, Pittsburgh, PA, USA.

Rui Qi Ji and Ervin Sejdic are with the Edward S. Rogers Department of Electrical and Computer Engineering, University of Toronto, Toronto, Ontario, Canada.

Ervin Sejdic is with the North York General Hospital, Toronto, Ontario, Canada.

Jessica K. Zègre-Hemsey is with the School of Nursing, University of North Carolina (UNC), Chapel Hill, NC, USA.

Salah Al-Zaiti is with the School of Nursing, University of Rochester, Rochester, NY, USA.

Salah Al-Zaiti is with the University of Rochester Medical Center (URMC), Rochester, NY, USA.

rapidly identify the less visually obvious ECG patterns linked to OMI.

The current state-of-the-art models for OMI classification are a random forest with engineered features and a proprietary convolutional neural network (CNN) [11][12]. One benefit of random forests is that they are easily explainable because they require the researcher to select the features to be used as input [2]. The features can be further interpreted using importance rankings, such as Shapley values. Furthermore, they do not require extremely large datasets to train. The downside of random forests is that they require multiple preprocessing steps to obtain the desired features. For ECG data, this may include filtering, median beat calculation, median beat segmentation, and feature calculation. Additionally, there may be useful information in the ECG that the selected features do not capture. Conversely, CNNs learn their own features automatically through the training process and therefore require fewer preprocessing steps [13]. This often leads to higher performance from CNNs than classical ML models like random forests [14]. The tradeoff is that CNNs require larger datasets to train and are less easily explainable since the model learns its own features.

ML models for ECG must be explainable for four main reasons [15][16]: 1) to justify clinical decisions made by the model; 2) to identify the shortcomings of the model in order to make improvements; 3) to identify new physiological patterns in the data; and 4) for users to understand when to override the decisions of the model.

To overcome the lack of inherent explainability of CNNs, researchers have used saliency maps to identify the most important regions of the input data for the model to make its prediction [17][18]. Within ECG research, saliency maps have often been used to confirm the clinical relevance of the features identified by CNN models [19][20][21][22][23][24].

The aim of this work is to design a clinically informed OMI classification model, ECG-SMART-NET, which can compete with the proprietary state-of-the-art CNN while maintaining a high level of clinical explainability [12]. OMI is linked with lead-specific temporal ECG features, including ST segment elevation [25][26]. OMI is also linked to cross-lead spatial ECG information such as reciprocal changes [25][26]. With these OMI features in mind, ECG-SMART-NET used a clinically informed design to first learn lead-specific temporal features followed by cross-lead spatial features. To maintain a high level of clinical explainability, saliency maps were leveraged, identifying the most important leads and within-lead regions for OMI classification. ECG-SMART-NET was compared to other CNN approaches and state-of-the-art random forest built using one of the largest prehospital ECG datasets with OMI labels.

## II. Methods

### A. Model Architecture

ECG-SMART-NET was based on the ResNet-18 architecture but required modifications. ResNet-18 includes convolutional and MaxPool layers with kernels of dimensions k x k and was designed for use with images of size 224 x 224 pixels. However, with 12-lead ECG data, these two-dimensional convolutional layers learn features across leads and time simultaneously. ECG data contains important lead-specific temporal features and cross-lead spatial features. Therefore, the traditional ResNet-18 architecture is not the most appropriate architecture for ECG data.

The inputs to the deep learning models in this work are 12-lead ECG median beats. It is important to first identify ECG components and their relevance to clinical practice before appreciating the architecture modifications that were made. Figure 1 shows the locations of the P wave, QRS complex, ST segment, and T wave on an ECG median beat with all 12 leads superimposed. Clinically relevant lead-specific temporal features can be calculated within each of these locations. These features, commonly linked to OMI, include ST segment elevation or depression and hyperacute T waves [25][26]. ST segment elevation occurs when the ST segment is above the isoelectric line. The isoelectric line marks no electrical activity. ST segment depression occurs when the ST segment is below the isoelectric line. Hyperacute T waves often have larger amplitudes, are broad-based, and are symmetrical. Hyperacute T waves can be a precursor to ST segment elevation.

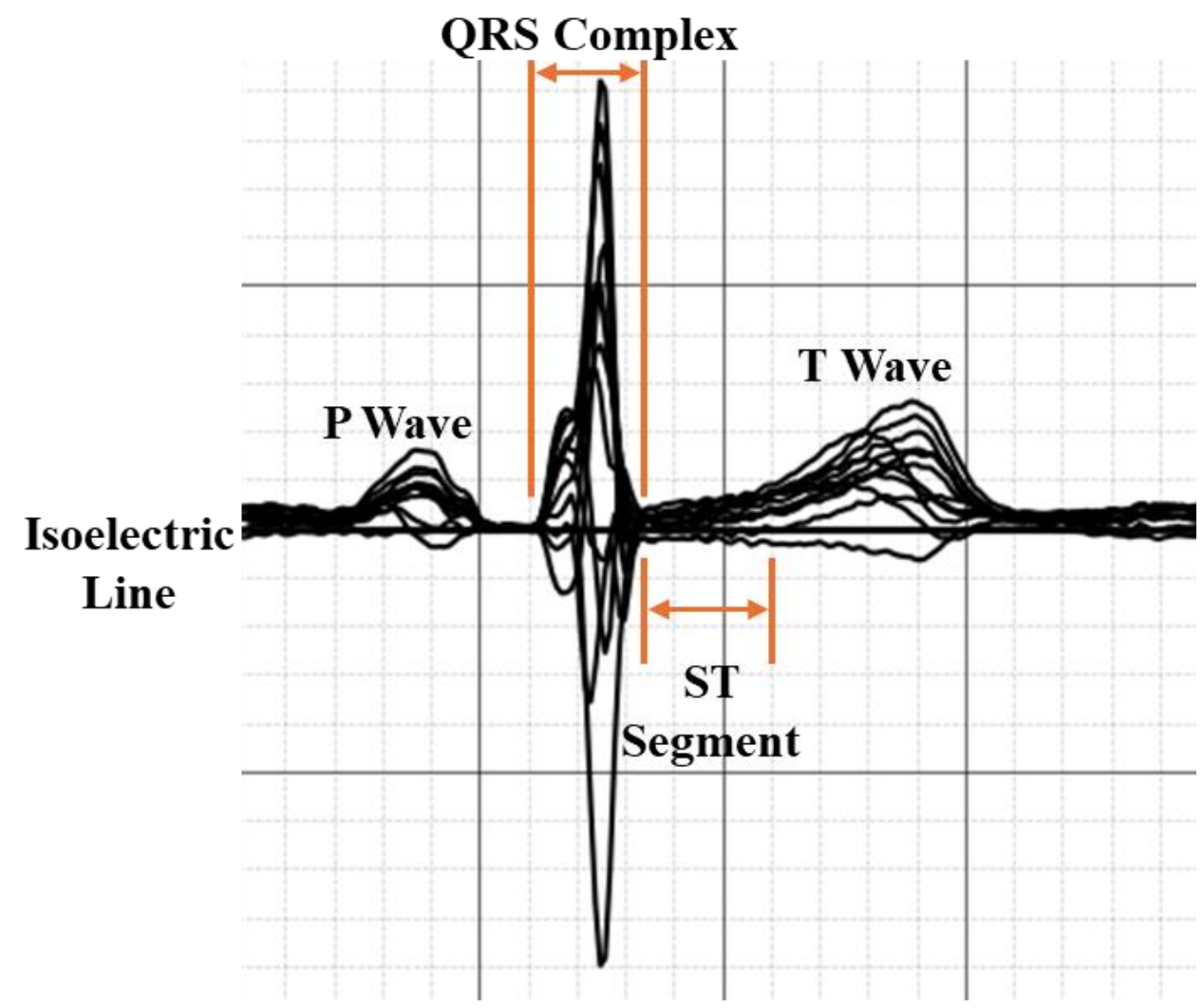

**Fig. 1.** This is an ECG median beat with all 12 leads overlayed. The important lead-specific temporal segments of this waveform are the P wave, QRS complex, ST segment, and T wave.

In addition to lead-specific temporal features, spatial features are important and are the reason multi-lead ECGs are recorded. Each ECG lead provides a different spatial view of a heartbeat. Figure 2 gives the locations of each of the 12 ECG leads in Cartesian space. Reciprocal changes are important spatial features for the identification of OMI [25][26]. A reciprocal lead is a lead opposite the direction of another lead. For example, lead aVL is a reciprocal lead of lead aVF and if changes are present in lead aVF, we should also look for changes in lead aVL. An example of an important reciprocal change is the presence of ST elevation in some leads, and ST depression in reciprocal leads, indicating potential STEMI.

Unlike ResNet-18, ECG-SMART-NET first identifies lead-specific temporal features and then learns cross-lead spatial features. This technique is similar to the approach used by a CNN for low ejection fraction [27]. Starting with the ResNet-18 architecture, all convolutional and MaxPool layers were modified to have kernels of dimensions 1 x k instead of k x k (temporal ResNet-18). This allowed the model to learn important temporal features within each lead of the ECG before collapsing across leads. To learn cross-lead spatial information, a spatial convolutional layer was added after the residual blocks with kernels of dimensions 12 x 1. Only one spatial convolutional layer was used to keep the model small as ECG data can be difficult to obtain. The ECG-SMART-NET architecture parameters are provided in Tables 1 and 2 for reproducibility. Table 1 gives the overall model architecture parameters and Table 2 provides the parameters for the residual blocks within the model.

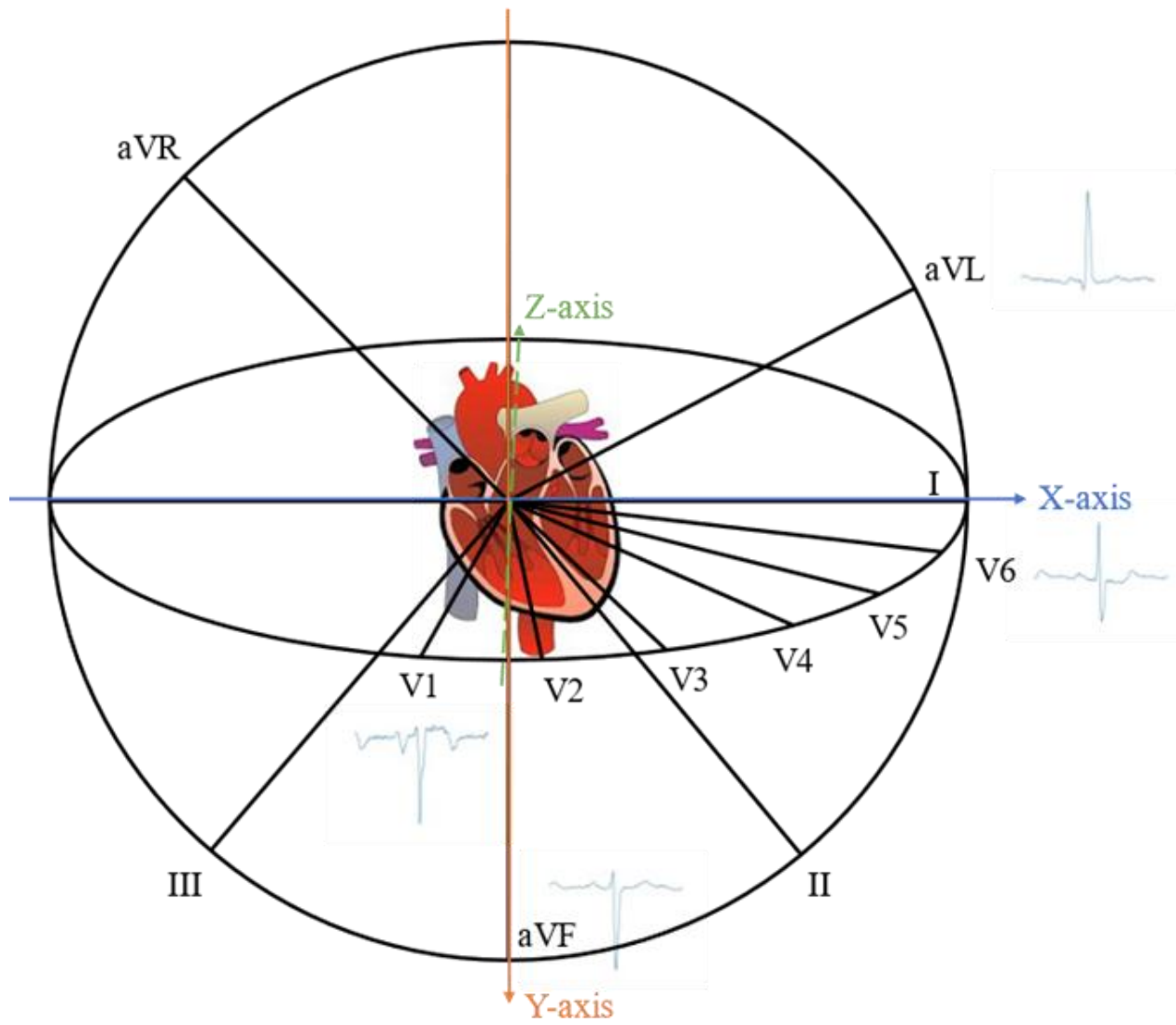

**Fig. 2.** The Cartesian coordinates for the 12 ECG leads.

TABLE I

ECG-SMART-NET MODEL ARCHITECTURE

| Layer | Output Dimensions | Parameters |
|---|---|---|
| Input | 1x12x200 | - |
| 2D Conv | 64x12x100 | Kernel: (1,7), Stride: (1,2), Padding: (0,3), Batch Norm, ReLU |
| MaxPool | 64x12x50 | Kernel: (1,3), Stride: (1,2), Padding: (0,1) |
| Residual Block 1 | 64x12x50 | First Stride1: (1, 1) |
| Residual Block 2 | 128x12x25 | First Stride2: (1, 2) |
| Residual Block 3 | 256x12x13 | First Stride3: (1, 2) |
| Residual Block 4 | 512x12x7 | First Stride4: (1, 2) |
| 2D Conv | 512x1x7 | Kernel: (12,1), Stride: (1,1), Padding: (0,0), Batch Norm, ReLU |
| AvgPool | 512x1x1 | (1,1) global average |
| Flatten | 512 | - |
| Fully Connected | 2 | (512, 2 classes) |

TABLE 2

ECG-SMART-NET RESIDUAL BLOCK

| Residual Block | |
|---|---|
| Layer | Parameters |
| 2D Conv | Kernel: (1,3), Stride: First Stride, Padding: (0,1), Batch Norm, ReLU |
| 2D Conv | Kernel: (1,3), Stride: (1,1), Padding: (0,1), Batch Norm |
| Shortcut Connection | ReLU |
| 2D Conv | Kernel: (1,3), Stride: (1,1), Padding: (0,1), Batch Norm, ReLU |
| 2D Conv | Kernel: (1,3), Stride: (1,1), Padding: (0,1), Batch Norm |
| Shortcut Connection | ReLU |

ECG-SMART-NET was compared to the original ResNet-18 model, a temporal ResNet-18 architecture that used 1 x k kernels, and current state-of-the-art random forest OMI model. The ResNet-18 model was pretrained on the ImageNet dataset [28]. The first convolutional layer of the pretrained ResNet-18 model was modified to take in one channel of data instead of 3 and the final fully connected layer was modified to have an output size of 2. The model architectures are provided in Figure 3.

The ResNet-18 architecture, in Figure 3C, was the original architecture that inspired both the temporal ResNet-18 and ECG-SMART-NET. ResNet-18 begins with a 2D convolution which learns information across leads and time simultaneously. The max pooling layer then downsamples the ECG in both spatial and temporal dimensions. The four residual blocks each consist of four convolutional layers. The last three residual blocks also downsample the ECG, using a stride of 2, so that larger features can be learned by the model. ResNet-18 then uses average pooling across leads and time so that we are left with only the channel dimension which is fed as features into a fully connected output layer. The ResNet-18 architecture collapses across leads very early on in the model, which does not give the model the opportunity to learn lead-specific features. Figure 3B shows the temporal ResNet-18 architecture. This model is identical to ResNet-18 except all kernels and strides are set to 1 in the lead dimension. This encourages the model to learn lead-specific features. The spatial dimension is only collapsed at the end of the model by the average pooling layer. Finally, Figure 3A shows the ECG-SMART-NET architecture. This model is the same as the temporal ResNet-18 except a spatial convolutional layer is added right after the residual blocks. After learning temporal features within each lead, the spatial convolutional layer learns information across all leads. This means that ECG-SMART-NET may learn cross-lead information that was missed by the temporal ResNet-18 model.

*B. Model Tuning*

The patients were split into train, validation, and test sets and were stratified by OMI labels so that OMI prevalence was maintained for each set. 80% of patients were kept in the training set and 10% were kept in each the validation and test sets. In the training set, serial ECGs were used. However, in the validation and test sets, all serial ECGs were removed. For both the validation and test sets only the first ECG acquired for each patient was kept. The splits are provided in Table 3.

TABLE 3

TRAINING/VALIDATION/TEST SPLIT

| Split | ECGs | Patients | OMI ECGs | OMI Patients |
|---|---|---|---|---|
| Training | 8912 | 5916 | 836 (9.4%) | 415 (7.0%) |
| Validation | 740 | 740 | 57 (7.7%) | 57 (7.7%) |
| Test | 741 | 741 | 62 (8.4%) | 62 (8.4%) |

The models were trained for up to 200 epochs with early stopping if validation loss did not improve after 10 consecutive epochs. The models utilized a cross-entropy loss function and an Adam optimizer with weight decay. To handle class imbalance, within each batch of training data, the majority class (no OMI) was randomly undersampled so that there was a 1:1 ratio of both classes, giving the model the chance to learn to identify the rare positive class. This method was implemented so new combinations of the majority class would be seen in each epoch. A grid search was run for each of the models to find the optimal weight decay (1e-2, 1e-3, 1e-4), batch size (32, 64, 128, 256), learning rate (1e-4, 1e-5, 1e-6), and learning rate for epoch 0 (1e-2, 1e-3, 1e-4). For ECG-SMART-NET the best hyperparameters were weight decay = 1e-3, batch size = 64, learning rate = 1e-4, and learning rate epoch 0 = 1e-4. For the temporal ResNet-18 the best hyperparameters were weight decay = 1e-3, batch size = 128, learning rate = 1e-4, and learning rate epoch 0 = 1e-4. For ResNet-18 the best hyperparameters were weight decay = 1e-2, batch size = 32, learning rate = 1e-4, and learning rate epoch 0 = 1e-2.

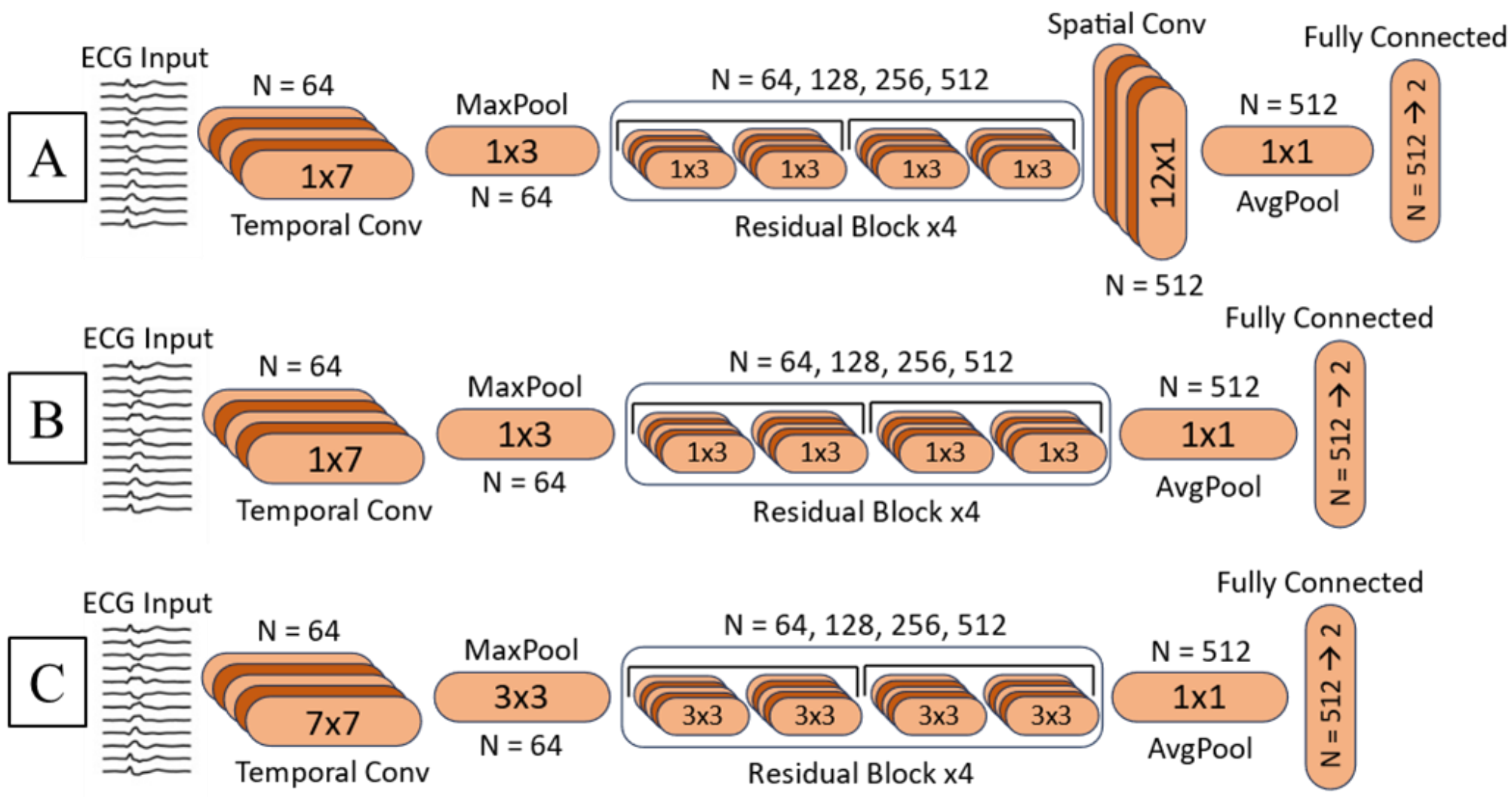

**Fig. 3.** Model Architectures: A) ECG-SMART-NET, B) Temporal ResNet-18, C) ResNet-18. These architectures are available at https://github.com/nateriek/ecg-smart-net.

The models were trained using PyTorch v2.4.0 and Python v3.11.9 in the Visual Studio Code Editor. The computer runs on the Windows 11 operating system with an Intel Core i7-9750H CPU, NVIDIA GeForce RTX 2080 GPU, and 32 GB RAM.

### *C. Saliency Maps*

Saliency maps highlight the most informative regions of a model's input that contribute to the model's outcome predictions. This can add significant explainability to a CNN. Saliency maps were generated using Grad-CAM in Python from the publicly available GitHub repository pytorch-grad-cam [29]. Grad-CAM highlights important regions of model input by using gradients of the target class with respect to feature maps in a selected convolutional layer. The saliency map is upscaled to the dimensions of the input data so that it can be overlaid as a heatmap [18]. We selected the positive OMI class as the target class and the final temporal convolutional layer to generate Grad-CAM heatmaps. These saliency maps provide insight into how different leads and ECG heartbeat segments influence the model's outcome prediction. A saliency map was calculated for all 62 positive OMI cases within the test set of the selected CNN model. Lead contributions were calculated by taking the sum of the saliency values within each lead and dividing by the total sum of saliency values across all leads. The saliency maps were mapped to a color scale ranging from blue (low importance) to red (high importance). Each saliency map was normalized between 0 and 1. Next, the 62 saliency maps were averaged together to generate one aggregate saliency map. This aggregate saliency map provides a clearer depiction of the leads and ECG segments that are most consistently contributing to OMI prediction. Finally, saliency maps were also averaged for LAD, LCX, and RCA occlusions separately to assess the saliency maps more locally. Grad-CAM was selected over gradient-based saliency maps because it utilizes higher-level feature maps from the last temporal convolutional layer, making it more robust to noise.

### *D. Dataset*

The dataset in this study consisted of prehospital 12-lead median beat ECGs from patients with a primary complaint of non-traumatic chest pain. The median beat ECG is a median of all beats from the original 10-second recording (500 s/s, 0.05-150 Hz). An example of both the 10-second and median beat ECGs are shown in Figure 4.

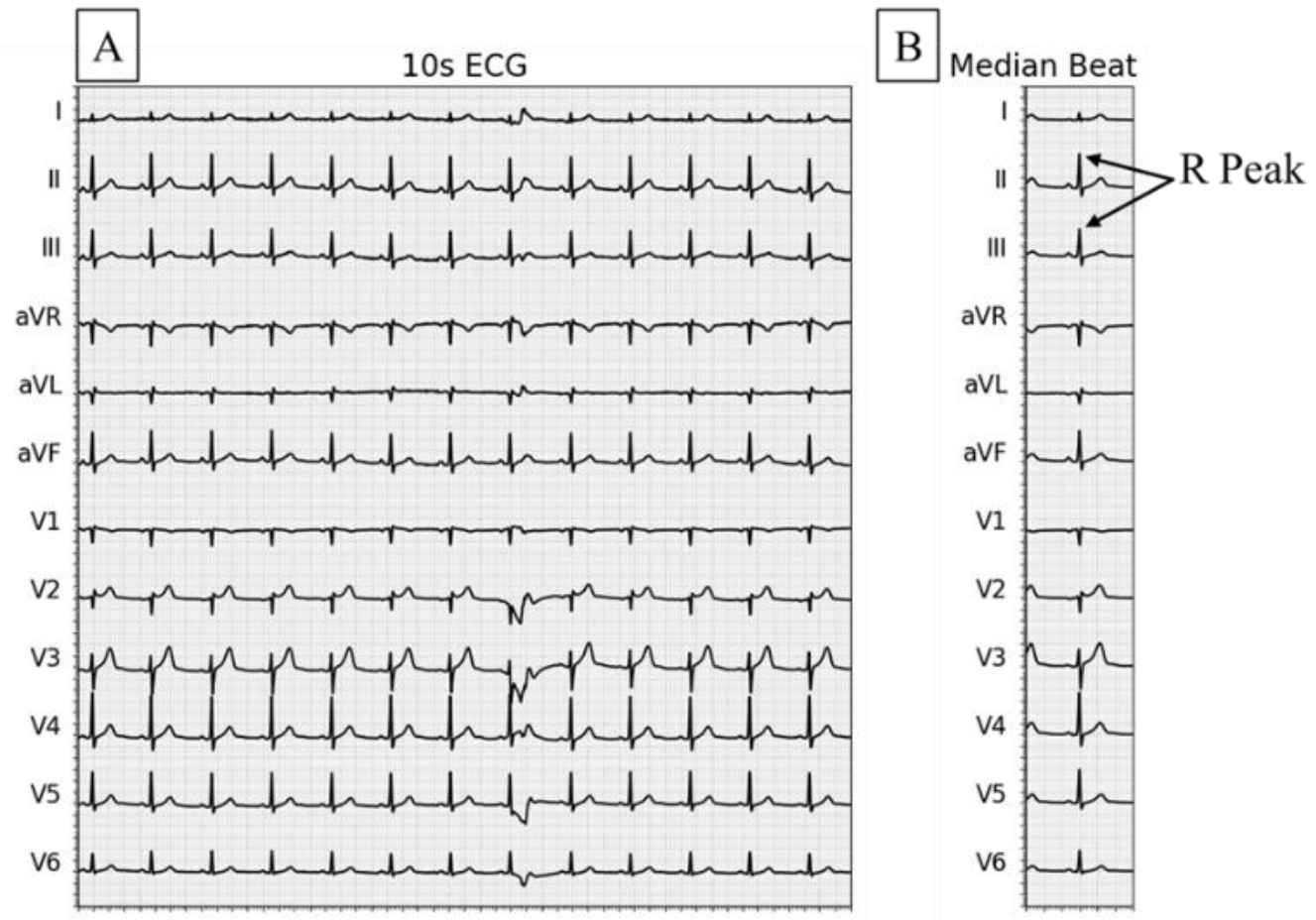

**Fig. 4.** (A) A raw 12-lead 10-second prehospital ECG recording. (B) The median beat ECG from the same recording, centered at the R Peak. Median beats were found by filtering the 10-second ECG, detecting beats centered at R peaks, removing ectopic beats, and taking the median of the remaining beats.

Data was collected from three EMS systems: one in Pittsburgh, PA (Pittsburgh EMS) and two in North Carlina (Orange County EMS and Mecklenburg County EMS). 11,309 prehospital ECGs were collected across all sites from patients 18 years or older with a chief complaint of non-traumatic chest pain. The median duration of symptoms onset to clinical presentation was 2.5 hours. 93 ECGs were excluded for ventricular tachycardia. 115 ECGs were excluded due to missing or noisy leads. Lastly, 708 serial ECGs were removed from the validation and test sets. A total of 10,393 ECGs from 7,397 unique patients with OMI labels were used in this study. Pittsburgh EMS and Mecklenburg County EMS used HeartStart MRX (Philips Healthcare) ECG machines while Orange County EMS used LIFEPAK 15 (Strkyer) ECG machines. OMI labels were created using angiographically defined culprit lesions. As mentioned in our previous work, OMI is defined as either 1) a TIMI flow of 0 or 1 in at least one of the three coronary arteries or their primary branches or 2) a TIMI flow of 2 with stenosis >70% and peak troponin of 5-10 ng/mL [11]. The patient cohort was 46% female with an average age of 59 and 7.2% OMI prevalence. The dataset is summarized in Table 4.

TABLE 4
DATASET DESCRIPTION

| | |
|---|---|
| ECGs | 10,393 |
| Patients | 7,397 |
| Age | 59 (±16) |
| Sex (Female) | 3392 (46%) |
| Race | |
| White | 3064 (41%) |
| Black | 2894 (39%) |
| Others | 93 (1.3%) |
| Unknown | 1346 (18%) |
| Ethnicity | |
| Not Hispanic | 5953 (80%) |
| Hispanic/Latino | 137 (1.9%) |
| Unknown | 1307 (18%) |
| ECG and Lab Findings | |
| Sinus Rhythm | 6029 (82%) |
| Atrial Fibrillation | 510 (6.9%) |
| Left BBB | 166 (2.2%) |
| Right BBB | 351 (4.7%) |
| ECG-LVH | 661 (8.9%) |
| Past Medical History | |
| Hypertension | 2870 (39%) |
| Diabetes | 1401 (19%) |
| High Cholesterol | 1882 (25%) |
| Current Smoker | 1181 (16%) |
| Known CAD | 1375 (19%) |
| Prior MI | 954 (13%) |
| Prior PCI | 143 (1.9%) |
| Prior CABG | 579 (7.8%) |
| Medical Therapy | |
| PCI | 689 (9.3%) |
| Total LAD Occlusion | 330 (4.5%) |
| Total LCX Occlusion | 247 (3.3%) |
| Total RCA Occlusion | 314 (4.2%) |
| Study Outcomes | |
| OMI | 534 (7.2%) |
| STEMI | 384 (5.2%) |

Abbreviations: bundle branch block (BBB), left ventricular hypertrophy (LVH), coronary artery disease (CAD), myocardial infarction (MI), percutaneous coronary intervention (PCI), coronary artery bypass surgery (CABG), left anterior descending artery (LAD), left circumflex artery (LCX), right coronary artery (RCA), occlusion myocardial infarction (OMI) ST-elevation myocardial infarction (STEMI).

The median beats have dimensions 12 leads x 600 datapoints sampled at 500 samples per second. Median beats were centered at the R Peak as shown in Figure 4. Before training any models, the median beats were shortened by removing the first 150 and last 50 points from each lead resulting in dimensions 12 leads x 400 datapoints. This was done to remove information from the prior and next beat so that the models would focus on the beat centered in the window. It is important to note that different windows may be used to accommodate longer QT intervals. For instance, the window can be adjusted to capture 400 milliseconds before and 800 milliseconds after the R Peak. The median beats were then downsampled to 250 s/s and were normalized by dividing each lead by its absolute maximum. This normalization was done to keep each ECG lead between -1 and 1 while preserving the isoelectric line. Normalizing within each lead allows the model to focus on waveform shape instead of relative amplitude differences. Also, it results in similar amplitude ranges for all leads so that the model can learn from all leads equally. Lastly, this normalization will not distort other leads if one lead has an abnormally high amplitude. The one negative of normalizing within each lead is relative amplitude information across leads is lost. ECG-SMART-NET was also evaluated on data normalized across leads between -1 and 1, but the validation performance was better when normalizing within each lead separately so within-lead normalization was used in our analyses. Median beats were calculated using our own algorithm. The steps of the algorithm are highlighted in Figure 5.

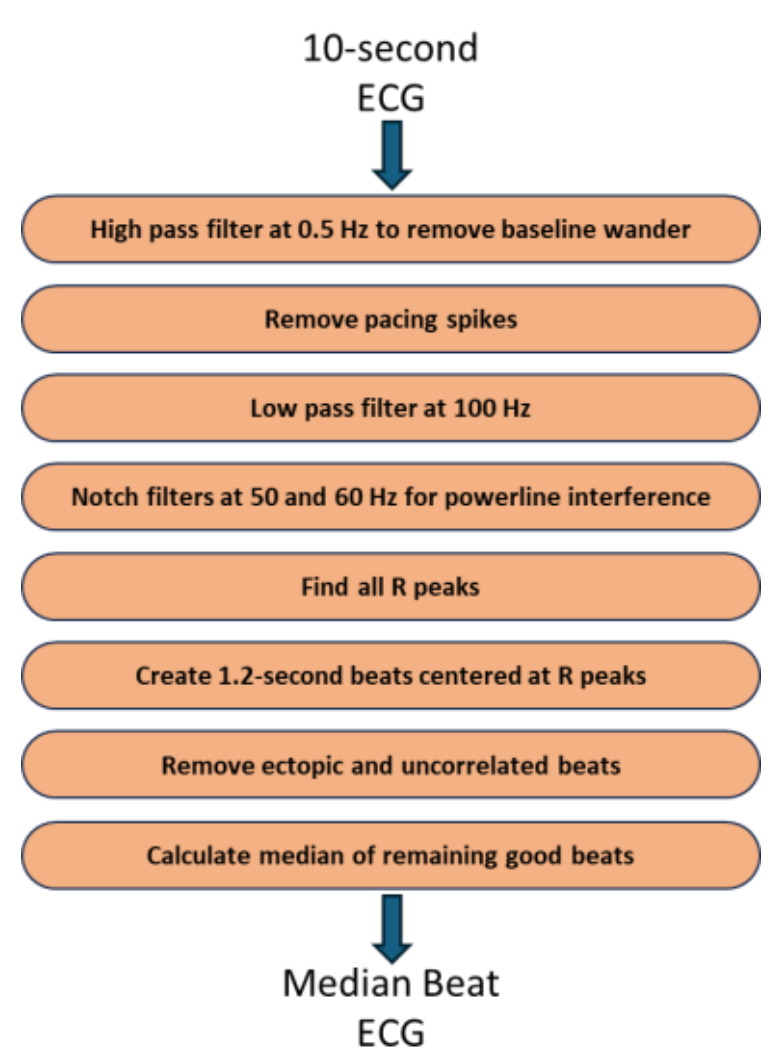

**Fig. 5.** Median beat calculation pipeline.

## III. Results

In section III, we first present a performance comparison between the CNN models in Figure 3 and the state-of-the-art random forest. The model performance comparisons are organized in Table 5 and Table 6 for the validation and test sets respectively. ECG-SMART-NET performance was then further examined with an ROC curve and precision-recall curve in Figure 5.

### *A. Model Performance*

Table 5 and Table 6 provide area under the receiver operating characteristics curve (AUC), average precision (AP), F1 score, sensitivity, specificity, positive predictive value, and negative predictive value on the validation and test sets respectively for ECG-SMART-NET and comparison models. AP is also sometimes referred to as the area under the precision-recall curve. An operating point was selected for each model by maximizing the F1 score on the validation set. The 95% confidence intervals for each metric, provided in brackets, were calculated using 1000 bootstrapped samples. This involved resampling the model predictions with replacement and computing the metrics for each bootstrap iteration. The two best performing models were ECG-SMART-NET and the state-of-the-art random forest. ECG-SMART-NET achieved the highest AUC and AP on the test set. The random forest achieved the second highest AUC and AP on the test set.

TABLE 5

VALIDATION MODEL PERFORMANCE METRICS

| Metric | ECG-SMART-NET | Random Forest | Temporal ResNet18 | Pretrained ResNet18 |
|---|---|---|---|---|
| Thresh | 0.91 | 0.47 | 0.90 | 0.82 |
| AUC | **0.943** [0.912, 0.971] | 0.942 [0.902, 0.975] | 0.935 [0.899, 0.967] | 0.937 [0.905, 0.968] |
| AP | 0.770 [0.679, 0.858] | **0.780** [0.682, 0.867] | 0.671 [0.552, 0.794] | 0.662 [0.537, 0.788] |
| F1 | **0.720** [0.622, 0.811] | 0.707 [0.596, 0.804] | 0.624 [0.521, 0.719] | 0.686 [0.578, 0.780] |
| Sens | 0.632 [0.510, 0.754] | 0.614 [0.489, 0.741] | **0.684** [0.554, 0.804] | 0.632 [0.500, 0.758] |
| Spec | **0.990** [0.981, 0.996] | **0.990** [0.981, 0.997] | 0.958 [0.942, 0.972] | 0.982 [0.972, 0.991] |
| Acc | **0.962** [0.949, 0.974] | 0.961 [0.947, 0.974] | 0.936 [0.919, 0.954] | 0.955 [0.941, 0.969] |
| PPV | **0.837** [0.723, 0.935] | 0.833 [0.714, 0.941] | 0.574 [0.459, 0.692] | 0.750 [0.622, 0.873] |
| NPV | 0.970 [0.957, 0.981] | 0.968 [0.955, 0.981] | **0.973** [0.961, 0.984] | 0.970 [0.955, 0.982] |

Abbreviations: area under the receiver operating characteristics curve (AUC), average precision (AP)

TABLE 6

TEST MODEL PERFORMANCE METRICS

| Metric | ECG-SMART-NET | Random Forest | Temporal ResNet18 | Pretrained ResNet18 |
|---|---|---|---|---|
| Thresh | 0.91 | 0.47 | 0.90 | 0.82 |
| AUC | **0.953** [0.921, 0.978] | 0.938 [0.895, 0.973] | 0.931 [0.890, 0.966] | 0.917 [0.864, 0.957] |
| AP | **0.843** [0.756,0.913] | 0.827 [0.739, 0.901] | 0.751 [0.647, 0.840] | 0.714 [0.597, 0.806] |
| F1 | 0.785 [0.690,0.862] | **0.792** [0.695, 0.870] | 0.711 [0.611, 0.800] | 0.678 [0.576, 0.765] |
| Sens | 0.677 [0.556, 0.791] | 0.677 [0.551, 0.789] | **0.694** [0.583, 0.806] | 0.661 [0.542, 0.772] |
| Spec | 0.996 [0.990, 1.000] | **0.997** [0.992, 1.000] | 0.976 [0.964, 0.987] | 0.973 [0.961, 0.985] |
| Acc | 0.969 [0.957, 0.980] | **0.970** [0.957, 0.982] | 0.953 [0.937, 0.968] | 0.947 [0.931, 0.962] |
| PPV | 0.933 [0.851, 1.000] | **0.955** [0.880, 1.000] | 0.729 [0.604, 0.839] | 0.695 [0.574, 0.811] |
| NPV | 0.971 [0.959, 0.983] | 0.971 [0.958, 0.983] | **0.972** [0.959, 0.984] | 0.969 [0.956, 0.981] |

Abbreviations: area under the receiver operating characteristics curve (AUC), average precision (AP)

ECG-SMART-NET achieved a validation AUC of 0.943, validation AP of 0.770, a test AUC of 0.953, and a test AP of 0.843. Figure 6 provides the ROC and Precision-Recall curves. The model performed slightly better on the test set than the validation set demonstrating good generalization to unseen data.

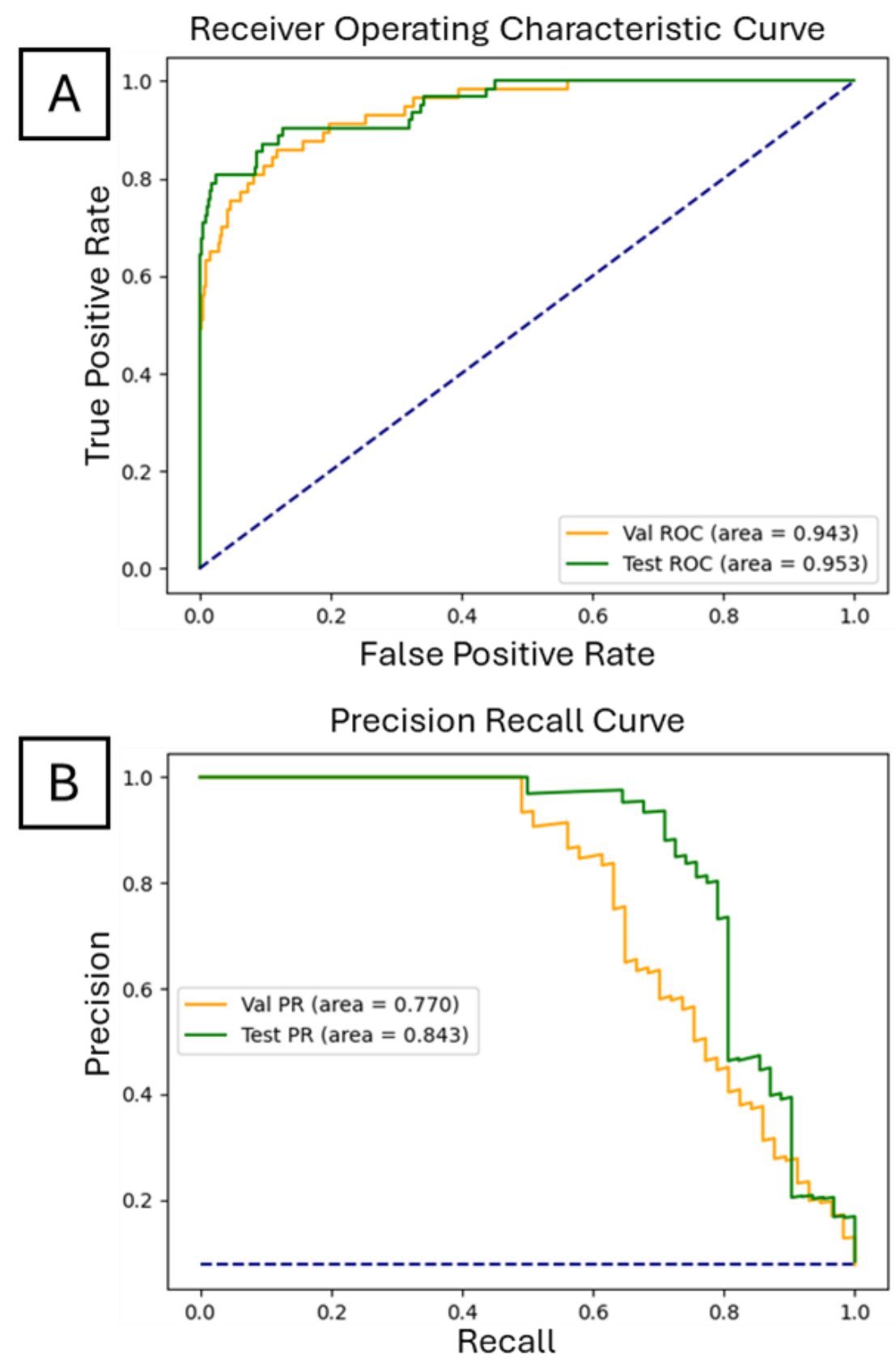

**Fig. 6.** ECG-SMART-NET model performance: A) ROC curve, B) Precision-Recall curve.

### *B. Saliency Map*

Figure 1 provides the different segments that make up a heartbeat and are visible on an ECG. These segments, in order, are the P wave, QRS complex, ST segment, and T wave. The aggregate saliency map across all 62 positive OMI cases in the test set is provided in Figure 7A. Figure 7B, 7C, and 7D contain the aggregate saliency maps for LAD, LCX, and RCA occlusions respectively. This figure gives insight into the temporal regions of an ECG heartbeat that most commonly help the model identify OMI. These regions, denoted by red, occur mainly in the terminal part of the QRS complex, ST segment, and early part of the T wave for most leads. Furthermore, the aggregate saliency map shows that leads aVL (lateral wall), lead III (inferior wall), and V2 (anterior wall) contribute most frequently to the classification of OMI in this specific dataset. It is important to note that saliency maps for individual ECGs show localized patterns of highest salience, so not all leads, nor the same group of leads, are important for every ECG.

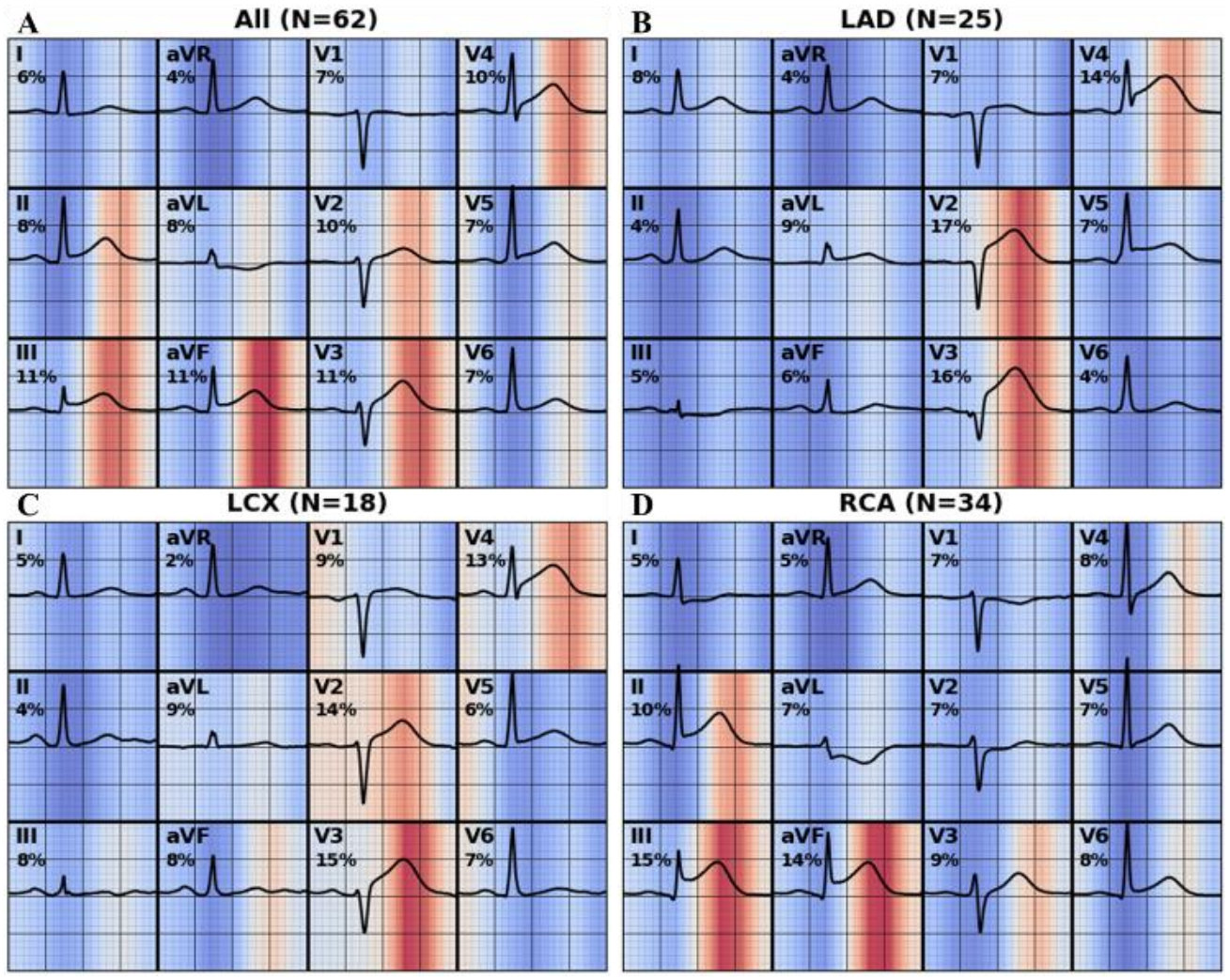

**Fig. 7.** Average saliency maps for A) All OMI (81% STEMI), B) LAD OMI (92% STEMI), C) LCX OMI (72% STEMI), D) RCA OMI (76% STEMI) cases in the test set.

### *C. Post-hoc Analyses*

Our preprocessing steps included downsampling from 500s/s to 250s/s and normalizing the median beat ECGs within each lead separately instead of across all leads. These steps were performed due to increases in performance on the validation data. We evaluated ECG-SMART-NET on median beats sampled at 500s/s and applied a lowpass filter at 150 Hz instead of 100 Hz. ECG-SMART-NET achieved a validation AUC of 0.936 and AP of 0.732. We also evaluated ECG-SMART-NET on median beats normalized across all leads and achieved a validation AUC of 0.936 and AP of 0.759. The selected method of downsampling to 250s/s and normalizing the median beat within each lead separately resulted in a validation AUC of 0.943 and AP of 0.770.

The test set includes 62 OMI patients with 50 of them having an ST-elevation myocardial infarction (STEMI) and 12 of them having a non-ST-elevation myocardial infarction (NSTEMI). To use ECG-SMART-NET as a screening tool for OMI, the decision threshold would be set to 0.53 (positive predictive value ≥ 35% on the validation set). With this threshold, 8 OMI positive ECGs were missed by ECG-SMART-NET (3 STEMI and 5 NSTEMI).

## IV. Discussion

We developed and evaluated a CNN based approach for OMI detection. ECG-SMART-NET captures temporal and spatial information from the ECG, mimicking a clinician. Furthermore, we leveraged skip connections to build a deeper model which can learn more complex features from the ECG. To our knowledge, this is the first CNN model that uses temporal followed by spatial convolution with skip connections for OMI detection. ECG-SMART-NET was inspired by ResNet-18 and was modified with clinical intuition to better suit the 12-lead ECG. We compared ECG-SMART-NET to both a pretrained ResNet-18 and temporal ResNet-18 and demonstrated superior performance in OMI detection. Furthermore, ECG-SMART-NET outperformed the state-of-the-art random forest on the test set. Superior model performance indicates that the ordering of temporal followed by spatial convolution provides a performance boost for ECG classification tasks. This makes sense clinically because clinicians identify temporal features within each of the ECG leads separately, and search for reciprocal changes across ECG leads.

The state-of-the-art CNN for OMI detection achieved a test AUC of 0.938, accuracy of 90.9%, sensitivity of 80.6%, and specificity of 93.7% on a different dataset [12]. ECG-SMART-NET achieved a test AUC of 0.953, accuracy of 96.9%, sensitivity of 67.7%, and specificity of 99.6%. Both models achieve high performance and strengthen the argument to use deep learning models for OMI prediction. However, the OMI prevalence in their test set was 21.6%, while the OMI prevalence in our test set was 7.2%, thus both models should be trained on a common dataset in the future.

ECG-SMART-NET outperforms the state-of-the-art random forest model on the test set and, with the use of saliency maps, can reduce concerns of explainability. Saliency maps allowed identification of the most important temporal features within each lead and the most important leads for OMI prediction for the selected test set of ECGs. The average saliency map showed that the most informative regions within each lead were the ST segment, and T wave. This is consistent with domain knowledge that ischemic injury causes electrical disturbances that affect both depolarization and repolarization. During ischemia, depolarization is slowed (reflected in the terminal QRS), while repolarization becomes heterogeneous (affecting the ST segment and T wave). These changes correspond to disruptions in signal propagation during phase 1 and phase 3 of the cardiac action potential [25][26]. The average saliency map also showed that lead aVL (lateral wall), lead III (inferior wall), and lead V2 (anterior wall) were the most informative for OMI classification. This is likely based on the distribution of occluded arteries (LCX, LAD, or RCA) in the selected test set and may vary for different datasets. These saliency maps enhance the explainability of ECG-SMART-NET, which is necessary for eventual clinical adoption. Although saliency maps can highlight important regions of the ECG for OMI detection, they do not tell us what about that region is important. Because of this, deep learning models remain less explainable than feature-based methods like a random forest model. There is a tradeoff between explainability and model performance. We have demonstrated that ECG-SMART-NET can identify OMI better than a traditional ResNet-18 and the current state-of-the-art random forest. Furthermore, we demonstrated the value of applying temporal convolutional layers followed by spatial convolutional layers for deep learning approaches with 12-lead ECG data. With saliency maps, we can add explainability to ECG-SMART-NET to identify the most informative leads and regions within lead for OMI classification. Another concern with saliency maps is the risk of confirmation bias. Researchers and clinicians may be tempted to interpret saliency maps in a way that confirms their expectations or clinical assumptions. However, saliency maps, and the underlying deep learning models, are imperfect and can highlight spurious or irrelevant ECG segments. Saliency maps should be used with caution and considered a supplementary resource rather than a definitive decision-making tool. Other multichannel biosignals, including electroencephalograms (EEG) or electromyograms (EMG), may benefit from ECG-SMART-NET for classification tasks due to its ability to learn both channel-specific temporal features and spatial features across channels.

Clinical implications for this work include enhanced identification of OMI using the 12-lead ECG. 12-lead ECG changes beyond ST segment elevation are not as visually obvious to clinicians in many cases. Leveraging explainable ML models that include the use of color-coded saliency maps will support objective, data driven, and timely decision support during moments of diagnostic uncertainty.

## V. Conclusion

We proposed a CNN, ECG-SMART-NET, for the prediction of OMI from 12-lead ECG data. The goal of this work was to identify if a clinically informed deep learning model could perform comparably or better than the state-of-the-art random forest on our dataset. On the test set, ECG-SMART-NET achieved an AUC of 0.953 and AP of 0.843. The random forest achieved an AUC of 0.938 and AP of 0.827 on the same test set. At the expense of some explainability, deep learning can outperform a feature-based classifier for OMI prediction. Explainability methods, such as saliency maps, can help narrow the gap in explainability between deep learning and feature-based methods. The use of either a deep learning model or a feature-based classifier should be determined based on the use case.